\documentclass[prb,aps,twocolumn,showpacs]{revtex4-2}

\usepackage{graphicx,color}
\usepackage{amsthm}
\usepackage{amsfonts}
\usepackage{algorithmic}
\usepackage{enumerate}
\usepackage{latexsym}
\usepackage{amsmath}
\usepackage{amssymb}
\usepackage{bm}
\usepackage[pdftex,plainpages=false,colorlinks=true,linkcolor=blue, citecolor=blue, urlcolor=blue]{hyperref}

\emergencystretch=\maxdimen
\begin{document}
\title{The spin Hall conductivity in the hole-doped bilayer Haldane-Hubbard model with odd-parity ALM}

\author{Minghuan Zeng$^{1}$}
\author{Ling Qin$^{2}$}
\author{Shiping Feng$^{3,4}$}
\author{Dong-Hui Xu$^{1,5}$}
\email{donghuixu@cqu.edu.cn}
\author{Rui Wang$^{1,5}$}
\email{rcwang@cqu.edu.cn}

\affiliation{$^{1}$Institute for Structure and Function \& Department of Physics \& Chongqing Key Laboratory for Strongly Coupled Physics, Chongqing University, Chongqing, 400044, P. R. China}

\affiliation{$^{2}$College of Physics and Engineering, Chengdu Normal University, Chengdu, 611130, Sichuan, China}

\affiliation{$^{3}$Department of Physics, Faculty of Arts and Science, Beijing Normal University, Zhuhai, 519087, China}

\affiliation{$^{4}$School of Physics and Astronomy, Beijing Normal University, Beijing, 100875, China}

\affiliation{$^{5}$Center of Quantum materials and devices, Chongqing University, Chongqing 400044, P. R. China}

\begin{abstract}
Spin current generated electrically is among the core phenomena of spintronics for driving high-performance spin device applications.
Here, on the basis of systematic investigations for the hole doped single-layer Haldane-Hubbard(HH) model,
we propose a new bilayer HH model to realize the compensated odd-parity spin splitting and the $T$-even spin
Hall conductivity where the two layers are connected by the time reversal transformation. Our results show that the vanishing layer-dependent electric potential
$V_{L}$ gives rise to odd-parity ALM protected by the combined symmetry $TM_{xy}$ with $T$ and $M_{xy}$ being the time reversal and mirror reflection perpendicular
to $z$ axis, and the $T$-even spin Hall conductivity simultaneously. In addition, though the staggered magnetization within each layer is substantially impacted by the
layer-dependent electric potential, small $V_{L}$'s only bring negligible changes to the net magnetization and the spin Hall conductivity,
indicating that the alternating spin splitting in momentum space and the spin Hall conductivity are insusceptible to external elements.
Most importantly, our work provides a general framework for the simultaneous realization of the compensated odd-parity spin splitting in momentum space
and the spin Hall conductivity in collinear magnets, in terms of stacked multi-layer systems.
\end{abstract}


\maketitle


{\it Introduction.}---The even-parity altermagnetism(ALM) induced by anisotropic electric crystal potential, characteristic of compensated and alternating spin polarization
in real and momentum space, has been intensively studied\cite{MaHY21,HuM25,Moreno12,Noda16,Okugawa18, Ahn19,Hayami19,Naka19,Libor20,Yuan20,Hayami20, Mazin21,Yuan21,Naka21,Rafael21,Shao21,Libor22_1,Libor22,Bai22,Bose22,Karube22,Dou25,Monkman25,Feng22,Shao22,Sun25}, including multiple time-reversal-symmetry(TRS)-breaking responses
such as the anomalous Hall effect\cite{Libor20,Feng22,Shao22}, charge to spin conversion\cite{Rafael21}, spin-splitter torque\cite{Rafael21,Karube22},
and giant tunneling magnetoresistance effects\cite{Libor22,Shao21,Sun25}.
Among these anomalous properties, the occurrence of spin current is at the core in this research field because of its important application in spintronics
such as the spin splitting torque\cite{Bai22,Karube22,Rafael21,Dou25}.

As the even-parity ALM and the associated physical properties are being investigated intensively\cite{Libor22_1,Libor22_2,Bai24},
the nonrelativistic odd-parity magnetism has recently attracted lots of research attention and become a rapidly developing research field\cite{Hellenes24,
Huang25,Song25,Brekke24,Ezawa25,Yu25,Lin25,Zeng25}. However, the odd-parity spin splitting observed earlier has been confined to
coplanar spin configurations\cite{Hellenes24,Song25,Yu25}, where the nonpreserving spin gives rise to short spin diffusion lengths\cite{Bai24}.
Until very recently, the sublattice current\cite{Lin25,Zeng25}, light\cite{Huang25,Zhu25,Liu25},
and orbit order\cite{Zhuang25} have been shown efficient for the occurrence of odd-parity ALM in the collinear antiferromagnet.
Most noteworthy, based on the spin group formalism\cite{Brinkman66,Litvin74,Litvin77}, we have recently demonstrated the sufficient condition
for the appearance of odd-parity ALM\cite{Zeng25}, the breaking nonmagnetic TRS,
and used the Haldane-Hubbard(HH) model\cite{Haldane88,He11} as an example to show that
it is the symmetry $T\bm{\tau}$ together with the breaking inversion symmetry of the Bravais lattice
in odd-parity ALM's that ensures the odd-parity spin splitting in momentum space. Here $T$ and $\bm{\tau}$ represent the usual time reversal
and the minimal real-space translation between two sublattices, respectively. However, after studying the transport properties of HH model,
we found that the symmetry $T\bm{\tau}$ prevents us from obtaining the spin-polarized current, which is of great importance in spintrinics.
In particular, it is essential to realize the spin-polarized current in odd-parity ALM's because the time reversal symmetry ensures the $T$-even spin-polarized
current that is analogous to the transversal spin current($T$ even) generated via the relativistic spin Hall effect (SHE) or/and the Rashba effect\cite{Miron11,Liu12_1,Liu12_2},
implying its important application in spintronics.

In this Letter, based on the symmetry analysis of the single-layer HH model with nonzero sublattice potentials\cite{He11,ifmmode16,Zheng15}[See also Sec. A2 in the supplemental material],
we find that at half-filling, instead of $T\bm{\tau}$ in the conventional HH model, the sublattices with opposite spin polarization are connected by
the combined symmetry of particle-hole transformation $\cal{P}$ and the inversion $\bar{E}$, i.e., $[C_2||{\cal{P}}\bar{E}]$ with $C_2$
being a 180$^\circ$ rotation around the axis perpendicular to spins, which ensures the compensated spin-splitting at half-filling.
In addition, the breaking symmetry $T\bm{\tau}$ indicates that $E_{\bm{k}\uparrow} \neq E_{-\bm{k}\downarrow}$, signaling the breakdown of the odd-parity ALM.
As the system deviates from half-filling, the symmetry $[C_2||{\cal{P}}\bar{E}]$ no longer holds, which leads to the existence of
nonzero net magnetization, then the traverse spin current appears due to the nonvanishing Berry curvature.
However, though the spin current is realized in the hole doped HH model with nonzero sublattice potentials, the characteristics of the odd-parity ALM
are destroyed by the breaking symmetry $T\bm{\tau}$. Therefore, we propose a new AA-stacking bilayer HH model with nonzero sublattice potentials[See Fig.\ref{Bilayer-HH-Model}]
to realize the compensated odd-parity spin splitting in momentum space and the $T$-even traverse spin conductivity simultaneously,
where the two layers are connected by the symmetry $TM_{xy}$ with $M_{xy}$ being the mirror reflection perpendicular to $z$ axis.
It is worth noting that this stacking scheme can be extended to the multiple-layer systems that have local and global $TM_{xy}$ symmetry, respectively, and the related work is ongoing.

\begin{figure}
\includegraphics[scale=0.25]{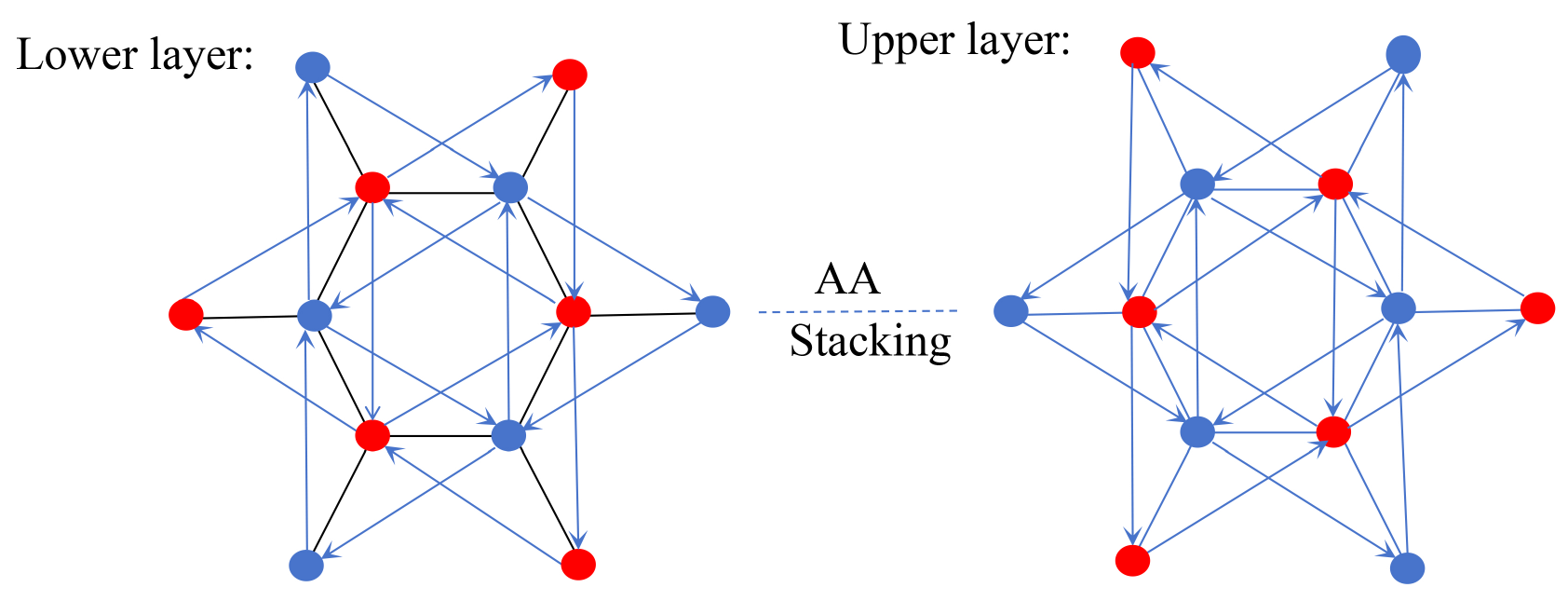}
\caption{(Color online) The schematic illustration of the AA-stacking bilayer HH model with the lower and upper layer plotted at the left and right,
respectively. Here the upward and downward spin polarizations are denoted by red and blue solid circles, and the arrows represent the sublattice currents.
We note that the spins in the first and second layer are polarized in the reversed direction, and sublattice currents flow
along opposite directions, reflecting the symmetry $TM_{xy}$ of this system.
\label{Bilayer-HH-Model}}
\end{figure}
{\it Theoretical model.}---Here we propose a new AA-stacking bilayer theoretical model consisting of two opposite single-layer HH model,
where the staggered magnetization and sublattice current are reversed from one layer to the other as illustrated in Fig.\ref{Bilayer-HH-Model}.
In addition, these two layers are coupled by the nearest-neighbor interlayer hopping processes.
Noteworthy, at the vanishing layer-dependent electric potential, the HH model at two layers are connected by the combined symmetry $TM_{xy}$,
which recovers the destroyed odd-parity ALM in the hole doped single-layer HH model with nonzero sublattice potentials.
This theoretical model not only preserves the $f$-wave ALM characteristics such as the alternate spin splitting in momentum space
and the compensated collinear magnetism in real space(between two layers), but also gives rise to the $T$-even spin Hall effect.
The above theoretical model can be explicitly expressed as ${\rm H} = \sum_{L=1,2} {\rm H}_{L} + {\rm H}_{12}$ with
\begin{subequations}\label{Bilayer-HH-Model}
\begin{eqnarray}
{\rm H}_{L} &=&-t\sum_{\langle ij\rangle}\big[{C_{iA}^{(L)}}^{\dagger}C_{jB}^{(L)}+\mathrm{H.C.} \big]
+ U\sum_{i,s=A,B}\hat{n}_{is\uparrow}^{(L)}\hat{n}_{is\downarrow}^{(L)} \nonumber\\
&+& \sum_{is}\big[ \mu+(-1)^s\frac{V_s}{2}+(-1)^{L-1}\frac{V_L}{2} \big]{C_{is}^{(L)}}^{\dagger}C_{is}^{(L)} \nonumber\\
&+& \lambda(-1)^{L-1}\sum_{s=A,B}\sum_{\langle\langle ij\rangle\rangle}{C_{is}^{(L)}}^{\dagger}e^{i\tfrac{\pi}{2}\nu_{ij}}C_{js}^{(L)}\;, \\
{\rm H}_{12} &=& -t_{\perp}\sum_{is}{C_{is}^{(1)}}^{\dagger}C_{is}^{(2)} + {\rm H.C.}\;,
\end{eqnarray}
\end{subequations}
where ${\rm H}_{L}$ and ${\rm H}_{12}$ represent the Hamiltonian of two layers and the nearest-neighbor hopping between them, respectively;
$\langle ij\rangle$ and $\langle\langle ij\rangle\rangle$ denote that the summation is over all the nearest- and next-nearest-neighbor sites, respectively;
${C_{is}^{(L)}}^{\dagger}=({C_{is\uparrow}^{(L)}}^{\dagger},{C_{is\downarrow}^{(L)}}^{\dagger})$ is a two-component spinor with $s=A,B$ denoting two sublattices;
$\nu_{ij}=\pm 1$ is the Haldane phase factor for clockwise and anticlockwise path connecting the next-nearest-neighbor sites $i$ and $j$;
$\hat{n}_{is\sigma}=C_{is\sigma}^{\dagger}C_{is\sigma}$ is the electron occupation number operator at site $is$ with spin $\sigma$.
$V_s$ and $V_L$ represent the sublattice potential and layer-dependent electric potential, respectively.
The detailed lattice set-up in real and momentum space is provided in Sec. I of the supplemental material.
In the following, the intralayer nearest-neighbor hopping integral $t$ and the lattice constant $a$ are set as the energy and length unit, respectively.

The above Hamiltonian is solved within the Hartree-Fock approximation with the onsite Coulomb interaction term being decomposed in an opposite manner,
\begin{eqnarray}
\hat{n}_{is\uparrow}^{(L)}\hat{n}_{is\downarrow}^{(L)} &\approx& \frac{n_{s}^{(L)}}{2}\sum_{\sigma}\hat{n}_{is\sigma}^{(L)}
- \sum_{\sigma}\sigma(-1)^{L+s-1}M_{s}^{(L)}\hat{n}_{is\sigma}^{(L)} \nonumber\\
&-& (\frac{{n_{s}^{(L)}}^2}{4}-{M_{s}^{(L)}}^2)\;,
\end{eqnarray}
where $n_{s}^{(L)}$ with $\sum_{Ls}n_{s}^{(L)}=4(1-\delta)$, $M_{s}^{(L)}$, and the chemical potential $\mu$ are self-consistently determined
for each set of onsite electron coulomb repulsion $U$, Haldane hopping $\lambda$, sublattice potential $V_{s}$, layer-dependent electric potential $V_{L}$,
and the hole doping $\delta$. The above mean-field parameters are explicitly calculated as
\begin{subequations}\label{MF-PARAM-1}
\begin{eqnarray}
n_{s}^{(L)} &=& \frac{1}{N}\sum_{i\sigma}\langle {C_{is\sigma}^{(L)}}^{\dagger}C_{is\sigma}^{(L)} \rangle \;, \\
M_{s}^{(L)} &=& \frac{(-1)^{L+s-1}}{2N}\sum_{i}\big[ \langle {C_{is\uparrow}^{(L)}}^{\dagger}C_{is\uparrow}^{(L)} \rangle
- \langle {C_{is\downarrow}^{(L)}}^{\dagger}C_{is\downarrow}^{(L)} \rangle  \big]\;, \nonumber\\
\end{eqnarray}
\end{subequations}
where $s=$ 0, 1 corresponds to sublattice A and B, respectively, and $N$ is the number of unit cells in the system.
The detailed derivation is provided in Sec. IB of the supplemental material.

\begin{figure}
\includegraphics[scale=0.42]{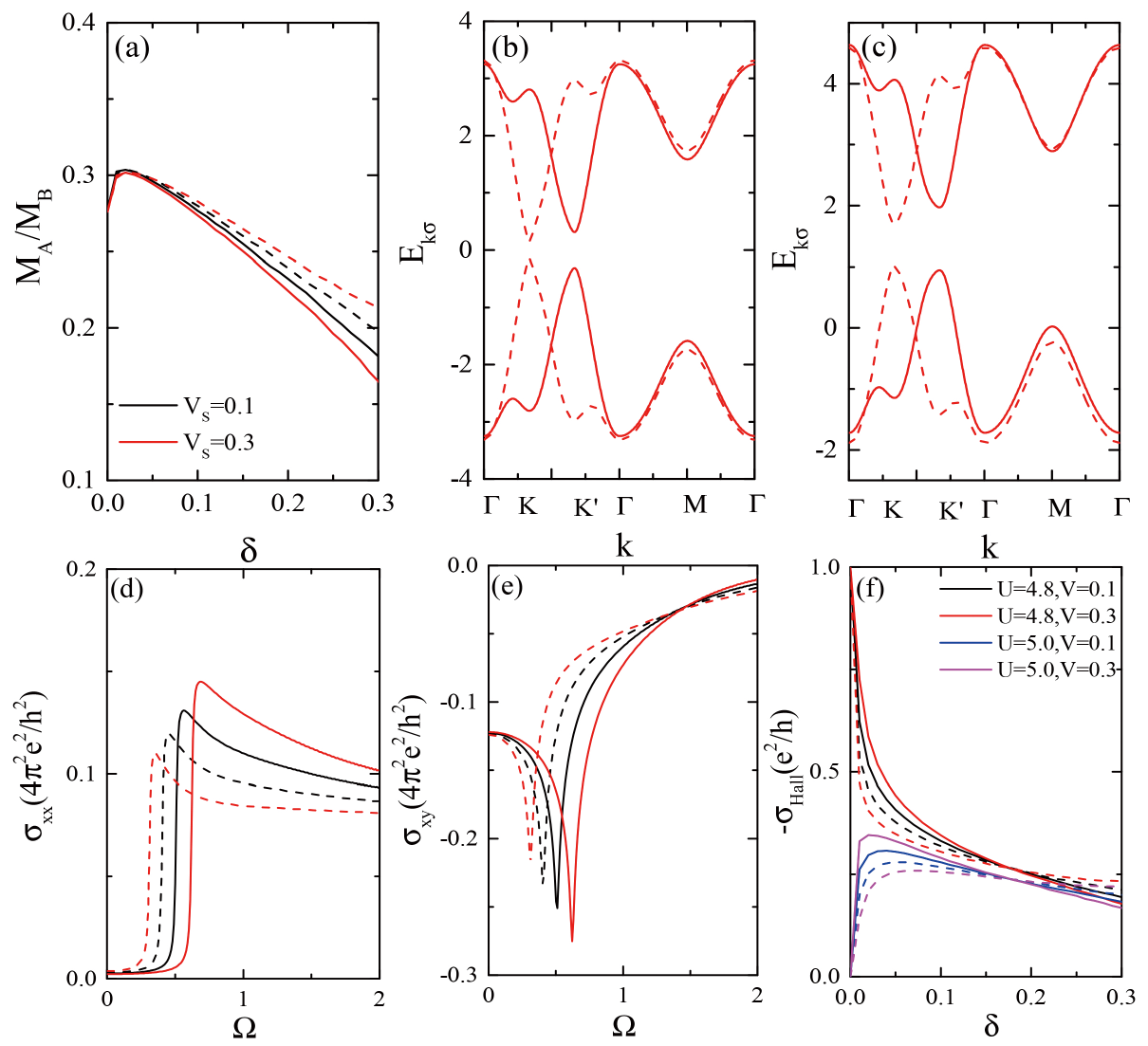}
\caption{(Color online) (a)The magnitude of staggered magnetization $M_{A}/M_{B}$ on sublattice A and B with up- and down-spin polarizations, respectively,
as a function of hole doping $\delta$ at $U=4.8$ and $\lambda=0.3$, where the black and red lines correspond to the sublattice
potential $V_s=$ 0.1 and 0.3, respectively. Here the solid and dashed lines correspond to up- and down-spin electron quasiparticles, and also applies to other figures.
The electron energy dispersion $E_{\bm{k}\sigma}$ as a function
of momentum in the zigzag direction at (b)$\delta=0$ and (c)$\delta=0.2$ for $U=4.8$, $\lambda=0.3$, and $V_{s}=0.3$.
The (d)longitudinal and (e)traverse optical conductivity, $\sigma_{xx}$ and $\sigma_{xy}$, coming from up- and down-spin electron quasiparticles
as a function of energy $\Omega$ at $\delta=0$ with $U=4.8$ and $\lambda=0.3$ for $V_{s}=$ 0.1(black) and 0.3(red);
(f)The up- and down-spin polarized anomalous Hall conductivity $\sigma_{\rm Hall}$ as a function of hole doping $\delta$
at $\lambda=0.3$ for $U=4.8$ with $V_{s}=$ 0.1(black), 0.3(red), and $U=5.0$ with $V_{s}=$ 0.1(blue), 0.3(magenta).
\label{Mag-E-EDisp-Cond-HallCD}}
\end{figure}
{\it Single-layer HH model with nonzero sublattice potentials}---Before discussing the spin splitting in momentum space and the spin Hall effect in the bilayer HH model\eqref{Bilayer-HH-Model},
we first study the hole doped single-layer HH model with nonzero sublattice potentials to illuminate the motivation for constructing the above bilayer HH model.
As pointed out in our previous work\cite{Zeng25}, the symmetry $T\bm{\tau}$ together with the breaking inversion symmetry of the Bravais lattice
in the odd-parity ALM enables the electron energy dispersion subject to the relation, $E_{\bm{k}\sigma}=E_{-\bm{k}-\sigma}$,
where $T$ and $\bm{\tau}$ represent the conventional time reversal and the minimal vector translation connecting
two sublatices, respectively. Thus to generate the spin-polarized current, the symmetry $T\bm{\tau}$ ought to be broken first.
For the collinear compensated magnetism in a bipartite lattice, the sublattice potential $V_s$ is the simplest way for the realization of the breaking symmetry of $T\bm{\tau}$.
The single-layer HH model with nonzero sublattice potentials is given explicitly in Eq.(1) of the supplemental material,
where the symmetry and the phase diagram of the half-filled system have been thoroughly investigated.
Here we first focus on the hole doping dependence of the magnitude of staggered magnetization on sublattice A($M_{A}$) and B($M_{B}$) at $U=4.8$ and $\lambda=0.3$.
As shown in Fig.\ref{Mag-E-EDisp-Cond-HallCD}(a), $M_{A}$ and $M_{B}$ exhibit a nonmonotonous behavior with the growth of hole doping $\delta$, implying that the system
at $U=4.8$ and $\lambda=0.3$, as well as $V_s \le 0.3$ is topologically nontrivial\cite{Zeng25}, coinciding with the nonzero Chern number with $C=2$.
More importantly, the separation between the staggered magnetization on sublattice A and B is enlarged monotonically as the system deviates from half-filling,
indicating that the HH model with nonzero sublattice potentials is spontaneously magnetized by the hole doping because of the breaking particle-hole symmetry $\cal{P}$,
and then the breaking combined symmetry $[C_2||{\cal{P}}\bar{E}]$.
In addition, the net magnetization $|M_{A}-M_{B}|$ is effectively strengthened when the sublattice potential is increased from $V_s=$ 0.1 to 0.3.

The electron energy dispersion $E_{\bm{k}\sigma}$ as a function of momentum in the zigzag direction at $\delta=0$ and $0.2$ for $U=4.8$
as well as $\lambda=0.3$ are plotted in Fig.\ref{Mag-E-EDisp-Cond-HallCD}(b) and (c), respectively, where the asymmetry between $E_{\bm{k}\downarrow}$(dashed lines) around valley $K$
and $E_{\bm{k}\uparrow}$(solid lines) around $K'$ reflects the breaking symmetry $T\bm{\tau}$ induced by sublattice potentials,
while the symmetric energy dispersion with respect to $E=0$ at half-filling corresponds to the combined symmetry ${\cal P}\bar{E}$[See Sec.IA2 of Sup. Mat.].
Interestingly, as shown in Fig.\ref{Mag-E-EDisp-Cond-HallCD}(c), for finite hole dopings, $E_{\bm{k}\sigma}$ around the valley $K$ and $K'$ exhibits an alternating behavior,
i.e., up- and down-spin electron energy dispersions are respectively centered around $K'$ and $K$, respectively,
while $E_{\bm{k}\sigma}$ around their midpoint $M$ is spin-polarized like the conventional ferromagnets. In this sense, at finite hole dopings, a novel magnetic state is identified
where the electron energy dispersion displays both the alternating and ferromagnetic
spin splitting in momentum space.

\begin{figure*}
\includegraphics[scale=0.3]{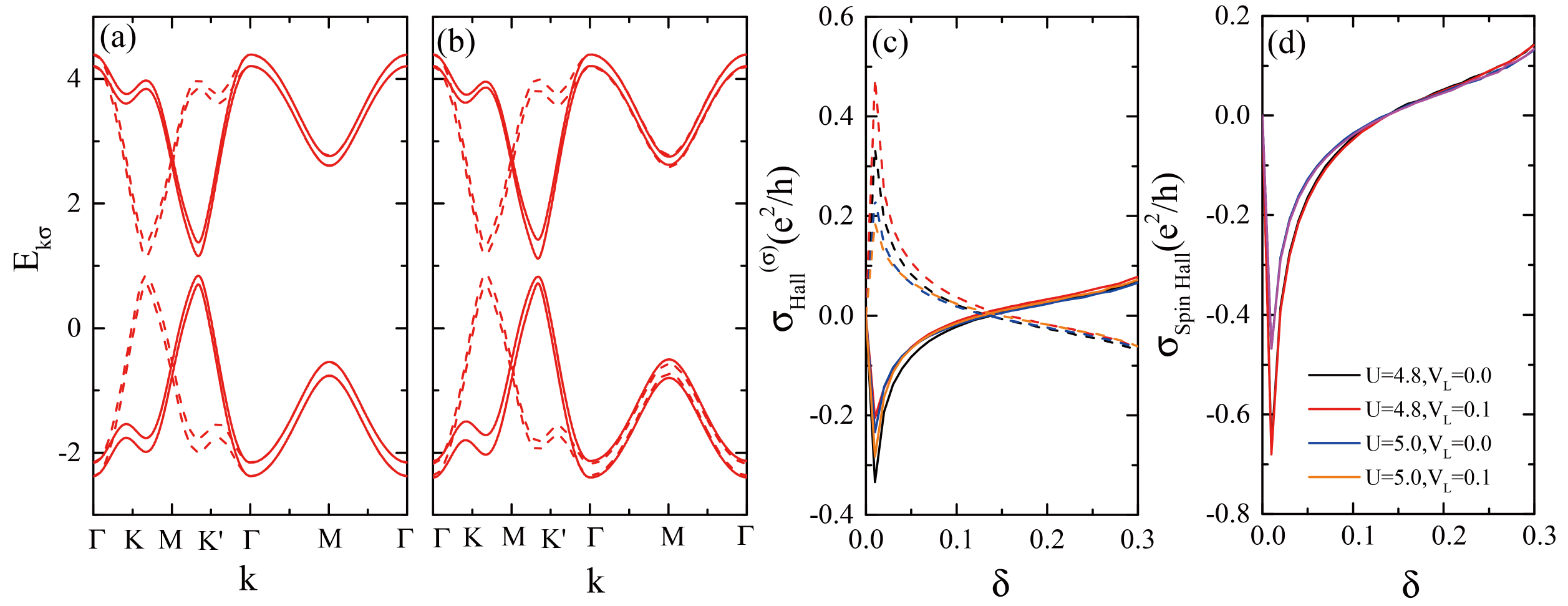}
\caption{(Color online) (a) The electron energy dispersion $E_{\bm{k}\sigma}$ as a function
of momentum in the zigzag direction at (a)$V_{L}=0$ and (b)$V_{L}=0.1$ for $U=4.8$, $\lambda=0.3$, $V_{s}=0.3$, $t_{\perp}=0.1$, and $\delta=0.1$.
Here the solid and dashed lines correspond to up- and down-spin electron quasiparticles, respectively, which also applies to other subfigures;
(c) The spin-dependent Hall conductivity $\sigma_{\rm Hall}^{(\sigma)}$ as a function of hole doping $\delta$ calculated from Eq.\eqref{Hall-Cond}
for $U=4.8$ as well as $V_L=$ 0(black), 0.1(red), and $U=5.0$ as well as $V_L=$ 0(blue), 0.1(magenta);
(d)The corresponding spin Hall conductivity $\sigma_{\rm spin \; Hall} = \sigma_{\rm Hall}^{(\uparrow)}-\sigma_{\rm Hall}^{(\downarrow)}$
as a function of $\delta$.
\label{EDispCut-HallCond-Bilayer-HH}}
\end{figure*}
Before discussing transport properties in the single-layer HH model with nonzero sublattice potentials, we first expand the electron energy dispersion[See Eq.(7) in the Sup. Mat.]
in the vicinity of $K$ and $K'$. The momentum around $K$ and $K'$ is calculated as $\bm{k} = K^{(')} + \bm{q}$, then the electron energy dispersion is approximated
as $E_{\bm{k}\sigma} \approx E_{\bm{q}s\sigma}= \mu + \tfrac{U}{2}(1-\delta) - \sigma\tfrac{U}{2}(M_{A}-M_{B})\pm \tfrac{1}{2}|\Delta_{s\sigma}|
\pm \tfrac{3}{4}\tfrac{|\bm{q}|^2}{|\Delta_{s\sigma}|}$ with $\Delta_{s\sigma}=V_{s}+\tfrac{U}{2}(n_{A}-n_{B})-s6\sqrt{3}\lambda-\sigma U(M_{A}+M_{B})$
where $s = \pm1$ holds for valley $K$ and $K'$, respectively. The inequivalence between $\Delta_{s\sigma}$ and $\Delta_{-s-\sigma}$ reflects a basic fact that the TRS
is broken by nonzero sublattice potentials, which then gives rise to the spin-dependent energy gap such that $\Delta_{-1\uparrow} \neq \Delta_{+1\downarrow}$.

As shown in Fig.\ref{Mag-E-EDisp-Cond-HallCD}(d) and (e), the optical conductivity $\sigma_{xx}$ and $\sigma_{xy}$ of the novel magnetic state at finite hole dopings
that is calculated from the Kubo formula\cite{Zeng25,Mahan90} display a qualitatively similar behavior to the system with vanishing $V_{s}$\cite{Zeng25},
i.e., are peaked in the vicinity of energy gap $\Delta_{s\sigma}$ because of the Van Hove singularity with $\nabla_{\bm{k}}E_{\bm{k}\sigma}|_{\bm{k}\to K,K'} = 0$.
Moreover, $\sigma_{xx}$ and $\sigma_{xy}$ are strongly spin-dependent in the presence of sublattice potential
that directly comes from the spin-dependent low-energy quasiparticles centered around $K$ and $K'$, respectively,
and the separation between up- and down-spin optical conductivities are monotonically enlarged by the increment of sublattice potential.
The spin-polarized Hall conductivity $\sigma_{\rm Hall}^{(\sigma)}$ calculated via Berry curvatures $B_{\sigma}^{(n)}(\bm{k})$\cite{Qiao10,Sato24,Zeng25}, i.e.,
\begin{eqnarray}\label{Hall-Cond}
\sigma_{\rm Hall}^{(\sigma)} = -\frac{e^2}{\hbar}\frac{1}{(2\pi)^2}\sum_{n}\int_{\rm BZ}d\bm{k}B_{\sigma}^{(n)}(\bm{k})n_{\rm F}(E_{\bm{k}\sigma}^{(n)}) \;,
\end{eqnarray}
is studied in Fig.\ref{Mag-E-EDisp-Cond-HallCD}(f). The results show that the spin-polarized currents appear as a direct consequence of the net magnetization
coming from the breaking symmetry $[C_2||{\cal{P}}\bar{E}]$ induced by hole doping at nonzero sublattice potentials, which further indicates that this spin
current is of odd parity with respect to the time reversal $T$.
In addition, the traverse spin current calculated via $\sigma_{\rm Hall}^{(\uparrow)} - \sigma_{\rm Hall}^{(\downarrow)}$ is significantly strengthened
by the increment of $V_{s}$ whether the system is topologically nontrivial or not, consistent with the enhanced net magnetization as the sublattice potential is escalated.

{\it Bilayer HH model with nonzero sublattice potentials}---The conventional compensated even-parity ALM's
break TRS with the electron energy dispersion $E_{\bm{k}\sigma} = E_{-\bm{k}\sigma}$,
and exhibit alternating spin splitting in momentum space determined by the rotation symmetry of the nonmagnetic state,
as well as the spin-polarized current that breaks TRS($T$-odd).
However, for compensated odd-parity ALM's, the symmetry $T\bm{\tau}$ prevents us from obtaining the spin-polarized current
though significant alternating spin splitting is observed in momentum space\cite{Zeng25}.
As shown above, the net magnetization and spin-polarized current occur at finite hole dopings in the single-layer HH model with nonzero sublattice potentials.
Thus we propose a bilayer HH model Eq.\eqref{Bilayer-HH-Model} with nonzero sublattice potentials to realize the compensated alternating spin splitting in momentum space
and the traverse spin current simultaneously. Here, as required by the combined symmetry $TM_{xy}$ with $M_{xy}$ being the mirror reflection
perpendicular to $z$ axis, the spin polarization and sublattice current are reversed from one layer to the other,
as shown in Fig.\ref{Bilayer-HH-Model}. Noteworthy, though the symmetry $T\bm{\tau}$ is broken by sublattice potentials for each layer,
the symmetry $TM_{xy}$ leads to the electron energy dispersion subjected to the relation $E_{(k_x,k_y)\sigma}=E_{(-k_x,-k_y)-\sigma}$.
On the other hand, the symmetry $TM_{xy}$ can be further broken by nonzero layer-dependent electric potentials $V_{L}$,
which then results in the negligible net magnetization proportional to $10^{-3}$ for small  $V_{L}$'s.

We in Fig.\ref{EDispCut-HallCond-Bilayer-HH}(a) and (b) study the electron energy dispersion $E_{\bm{k}\sigma}$ as a function
of momentum in the zigzag direction for $V_{L}=0$ and 0.1, respectively, as well as $U=4.8$, $\lambda=0.3$, $V_{s}=0.3$, $t_{\perp}=0.1$, and $\delta=0.1$,
where the solid and dashed lines describe $E_{\bm{k}\sigma}$ for up- and down-spin electron quasiparticles.
The electron energy dispersion displays the compensated alternating spin splitting in the vicinity of valley $K$ and $K'$,
while the TRS character of electron energy dispersions at $V_{L}=0$ is broken by nonzero layer potentials. Interestingly, though $E_{\bm{k}\sigma}$
exhibits spin splitting around the midpoint $M$ between $K$ and $K'$, its magnitude is significantly smaller than the single-layer system with nonzero sublattice potentials
and hole dopings, indicating that the magnetic properties of the bilayer HH model are less susceptible to external elements.
Then the spin-dependent Hall conductivity $\sigma_{\rm Hall}^{(\sigma)}$ as a function of hole doping is studied at $\lambda=0.3$, $V_{s}=0.3$, $t_{\perp}=0.1$,
as well as $U=4.8$ and $V_{L}=0$(black), $U=4.8$ and $V_{L}=0.1$(red),
$U=5.0$ and $V_{L}=0$(blue), $U=5.0$ and $V_{L}=0.1$(magenta), where the solid and dashed lines represent up- and down-spin polarized Hall conductivities, respectively.
The results show that the Hall conductivity is strongly spin-polarized at finite hole dopings, especially that the up- and down-spin polarized $\sigma_{\rm Hall}^{(\sigma)}$
propagate in opposite directions, which then naturally gives rise to the spin Hall effect. We emphasize that this spin Hall effect directly
arises from the opposing net magnetization and Berry curvatures in the two layers.
Most noteworthy, as shown in Fig.\ref{EDispCut-HallCond-Bilayer-HH}(d), though the layer potential $V_{L}$ brings substantial impacts on spin-polarized
Hall conductivity, the net spin Hall conductivity $\sigma_{\rm Spin\; Hall}=\sigma_{\rm Hall}^{(\uparrow)}-\sigma_{\rm Hall}^{(\downarrow)}$ coming from nonzero Berry curvatures
is insusceptible to nonzero $V_{L}$'s. In addition, whether the single-layer HH model is topologically nontrivial($U=4.8$, $\lambda=0.3$) or not($U=5.0$, $\lambda=0.3$),
$\sigma_{\rm Spin\; Hall}$ exhibits a nonmonotonous behavior, i.e., first surges as the system deviates from half-filling, then is peaked at $\delta\approx 0.01$, and
gradually decreases with the further increment of hole doping.

{\it Discussion and conclusion}---After the compensated odd-parity spin splitting was realized by introducing the sublattice current\cite{Lin25,Zeng25},
light\cite{Huang25,Zhu25,Liu25}, or orbit order\cite{Zhuang25} to break the nonmagnetic TRS, this odd-parity spin splitting
has been proved as the odd-parity ALM using the spin group method\cite{Zeng25}.
However, the symmetry $[C_2||\bar{E}]$ in the nontrivial spin space group that corresponds to the symmetry $T\bm{\tau}$ prevents us from obtaining the spin-polarized current.
In this work, we first show that a novel magnetic state exists in the HH model with nonzero sublattice potential at finite hole dopings, where the net magnetization and
spin-polarized current appear. Then a new bilayer HH model is proposed to realize the simultaneous occurrence of odd-parity spin splitting
and traverse spin current where the staggered magnetization and Haldane hopping are reversed between two layers while the sublattice potential keeps identical,
indicating that the two layers are connected by the time reversal transformation. Our results show that the symmetry $TM_{xy}$ of the bilayer HH model
with vanishing layer potential $V_{L}$ ensures the odd-parity ALM, while the opposite up- and down-spin polarized traverse conductivity in the lower and upper layer,
respectively, gives rise to the substantial spin Hall conductivity. In addition, the net magnetization brought by nonzero $V_{L}$'s that break the symmetry $TM_{xy}$ is negligible
which is proportional to $10^{-3}$ for $V_{L}=0.1$, and nonzero $V_{L}$'s only bring small quantitative changes to the spin Hall conductivity.
Most importantly, the proposal of the bilayer HH model for the realization of odd-parity ALM and spin Hall effect can be naturally extended to other systems
with different symmetry, thus our work brings about a general framework for the generation of spin-polarized currents in collinear odd-parity ALM's.

\begin{acknowledgments}

Minghuan Zeng thanks Y.-J. Wang for fruitful discussions.
This work was supported by the National Key Re
search and Development Program of China under Grant
 Nos. 2023YFA1406500 and 2021YFA1401803, the Na
tional Natural Science Foundation of China (NSFC)
 under Grant Nos. 12222402, 92365101, 12474151,
 12347101, 12247116, 12274036, and 12404121, Beijing National
 Laboratory for Condensed Matter Physics under Grant
 No. 2024BNLCMPKF025, the Fundamental Research
 Funds for the Central Universities under Grant No.
 2024IAIS-ZX002, the Chongqing Natural Science Foun
dation under Grants Nos. CSTB2023NSCQ-JQX0024
 and CSTB2022NSCQ-MSX0568, and the Special Fund
ing for Postdoctoral Research Projects in Chongqing
under Grant No. 2024CQBSHTB3156.

\end{acknowledgments}

\bibliography{BIBHH-Model}

\end{document}